\documentclass[11pt]{article}

\usepackage{amssymb,amsmath}
\usepackage[dvips]{graphicx}
\usepackage{verbatim}

\DeclareMathOperator{\poly}{poly}
\DeclareMathOperator{\Tr}{tr}

\newtheorem{thm}{Theorem}

\newtheorem{cor}[thm]{Corollary}
\newcommand{\vskipline}{\vskip 11pt}

\newcommand{\set}[1]{\lbrace #1 \rbrace}

\newcommand{\bra}[1]{\langle #1 |}
\newcommand{\ket}[1]{| #1 \rangle}

\newcommand{\norm}[1]{\lVert #1 \rVert}
\newcommand{\tensor}{\otimes}
\newcommand{\Tensor}{\bigotimes}

\newcommand{\Hil}{\mathcal{H}}
\newcommand{\RR}{\mathbb{R}}

\newcommand{\SSS}{\mathcal{S}}

\begin{document}

\title{The Local Consistency Problem for Stoquastic and 1-D Quantum Systems}

\author{Yi-Kai Liu\\
Institute for Quantum Information\\
California Institute of Technology\\
\texttt{yikailiu@caltech.edu}}

\date{}

\maketitle


\begin{abstract}
The Local Hamiltonian problem (finding the ground state energy of a quantum system) is known to be QMA-complete.  The Local Consistency problem (deciding whether descriptions of small pieces of a quantum system are consistent) is also known to be QMA-complete.  Here we consider special cases of Local Hamiltonian, for ``stoquastic'' and 1-dimensional systems, that seem to be strictly easier than QMA.  We show that there exist analogous special cases of Local Consistency, that have equivalent complexity (up to poly-time oracle reductions).  Our main technical tool is a new reduction from Local Consistency to Local Hamiltonian, using SDP duality.
\end{abstract}



\section{Introduction}

Local Hamiltonian is the problem of estimating the ground state energy of a quantum system with local interactions.  This is an important problem in condensed matter physics and quantum chemistry, and it is an interesting ``matrix-valued'' generalization of classical problems like Max-k-SAT.  Local Hamiltonian also plays a significant role in complexity theory:  it is QMA-complete, where QMA (Quantum Merlin-Arthur) is a generalization of NP where one allows the witness to be a quantum state \cite{KSV,AN,KKR,OT}.  

Two special cases of Local Hamiltonian are of particular interest, because they illustrate the differences in complexity between ``quantum'' and ``classical'' problems.  First, consider a 1-D system:  a collection of $d$-dimensional particles (qudits) arranged on a line, with nearest neighbor interactions.  Such systems have been extensively studied in condensed matter physics, and in many cases they can be solved efficiently using heuristic methods (see, e.g., \cite{vidal-1d,schollwock}).  On the other hand, this problem was recently shown to be QMA-hard for $d \geq 12$ \cite{qma1d-1,qma1d-2}, and its complexity for smaller $d$ is still open.  This complicated picture is very different from the classical situation:  when restricted to a 1-D chain, nearly all classical constraint satisfaction problems can be solved exactly in polynomial time, using dynamic programming.  The reason seems to be that 1-D quantum systems can have much a richer correlation structure, due to quantum entanglement, than 1-D classical systems.

Secondly, consider a ``stoquastic'' quantum system \cite{stoq-1,stoq-2}.  Here, the Hamiltonian has a certain generic form, which ensures that the ground state is a superposition of classical basis states with real non-negative coefficients; thus it more closely resembles a classical probability distribution.  (This feature is found in many physical systems, as well as some quantum algorithms \cite{FGGS}.)  Stoquastic Local Hamiltonian is in AM, which suggests that it is strictly easier than QMA (unless QMA is in AM, which would be somewhat surprising); however, it is also MA-hard, hence it is at least as hard as classical NP-complete problems \cite{stoq-1}.  Also, Stoquastic Local Hamiltonian is complete for a peculiar class, StoqMA \cite{stoq-2}.  Thus stoquastic systems occupy a curious middle ground between the quantum and classical regimes.

In this paper we approach these questions from a different direction.  We consider the Local Consistency problem, also called consistency of density matrices \cite{A}:  given several descriptions of small pieces of a quantum system, decide whether they are consistent with a single overall state.  This problem is QMA-complete under poly-time oracle reductions \cite{Liu-consistency-qma}.  (A related problem in quantum chemistry, $N$-representability, is also QMA-complete \cite{Liu-N-rep}.)  Thus, the Local Consistency and Local Hamiltonian problems have equivalent complexity, up to poly-time oracle reductions.  In this paper we show that this equivalence also holds for special cases involving 1-D and stoquastic systems, which are not known to be QMA-hard.  

Our results are as follows.  We define special versions of the Local Consistency problem for 1-D and stoquastic systems, and, in these special cases, we give poly-time oracle reductions from Local Hamiltonian to Local Consistency and vice versa.  The reductions in the forward direction are similar to \cite{Liu-consistency-qma}, but the backwards reductions use a new idea based on semidefinite programming duality.  (Note that, since these problems are not known to be QMA-hard, we cannot use the machinery of QMA-completeness; instead we have to show reductions in both directions explicitly.)  

This duality idea is interesting in its own right, as it shows a basic connection between the Local Hamiltonian and Local Consistency problems.  Specifically, we show a poly-time oracle reduction from Local Consistency to Local Hamiltonian, that preserves the structure of the quantum system (i.e., the number of qubits and the subsets of qubits which interact).  As mentioned earlier, this reduction also works in cases that are not QMA-hard.  

This is quite different from the reduction that is obtained using QMA-completeness.  There, one gets a poly-time mapping reduction, from Local Consistency to a version of Local Hamiltonian that is QMA-hard.  That reduction does not preserve the structure of the quantum system; the reason is that the Hamiltonian works by ``simulating'' the QMA verifier for Local Consistency, which requires many additional qubits.  (In particular, the witness for Local Consistency contains multiple copies of the original system \cite{Liu-consistency-qma}, and additional qubits are used to simulate the verifier's computation \cite{KSV,AN}.)  

Independently, a similar duality technique was used in \cite{Hall} to study a related problem, known as ``subsystem compatibility.''  There, one is given descriptions of large subsets of the system, i.e., all but one of the qubits.  For a system of $n$ qubits, the input is exponentially large in $n$, and one can solve the problem in time polynomial in the length of the input.  In our problem, Local Consistency, one is given descriptions of small subsets, i.e., $k$ qubits, where $k = O(1)$.  Then the input is only polynomially large in $n$, and in polynomial time, one gets a reduction to Local Hamiltonian instead.

Outline of the paper:  In section 2 we review the definitions of the problems and the basic tools used in the reductions.  In section 3 we present our new reduction from Local Consistency to Local Hamiltonian via SDP duality.  In section 4 we apply this to 1-D systems, and in section 5 we apply this to stoquastic systems.

\section{Preliminaries}

\subsection{Definitions of Problems}

We briefly review some basic definitions regarding quantum states; see \cite{NC} for a thorough introduction.  Suppose we are interested in a system of $n$ qubits.  This is associated with a vector space $\Hil = \mathbb{C}^{2^n}$.  $\Hil$ has a tensor product structure, $\Hil = \mathbb{C}^2 \tensor \cdots \tensor \mathbb{C}^2$ ($n$ times), and an orthonormal basis consisting of the vectors $\ket{z} = \ket{z_1} \tensor \cdots \tensor \ket{z_n}$, where $z_1,\ldots,z_n \in \set{0,1}$ (these correspond to the classical states of the system).  

An operator on the system is a matrix acting on $\Hil$.  An operator on a subset of qubits $C \subseteq \set{1,\ldots,n}$ is a matrix of the form $M \tensor I$, where $M$ acts on the tensor factors of $\Hil$ corresponding to $C$, and $I$ acts on the tensor factors corresponding to $\set{1,\ldots,n}-C$.  

A quantum state is represented by a density matrix $\rho$, which is a positive semidefinite matrix with trace 1 acting on $\Hil$.  A subset of qubits $C$ is described by a reduced density matrix $\Tr_{\set{1,\ldots,n}-C}(\rho)$ (called the partial trace of $\rho$); this is analogous to computing a marginal probability distribution by integrating over the unwanted variables.

\vskipline

The Local Hamiltonian problem is defined as follows \cite{KSV,AN}:  
\begin{quote}
Consider a system of $n$ qubits.  We are given a Hamiltonian $H = H_1+\cdots+H_m$, where each $H_i$ acts on a subset of qubits $C_i \subseteq \set{1,\ldots,n}$ (and so has dimension $2^{|C_i|} \times 2^{|C_i|}$).  The $H_i$ are Hermitian matrices, with norm $\norm{H_i} \leq 1$.  Also, each subset $C_i$ has size $|C_i| \leq k$, for some fixed $k$.  

In addition, we are given a string ``$1^s$'' (the unary encoding of a natural number $s$), and two real numbers $a$ and $b$, such that $b-a \geq 1/s$.  

All numbers are specified with $\poly(m,s)$ bits of precision.  

The problem is to distinguish between the following two cases:  
\begin{itemize}
\item If $H$ has an eigenvalue that is $\leq a$, output ``YES.''  
\item If all the eigenvalues of $H$ are $\geq b$, output ``NO.''  
\end{itemize}
\end{quote}

Note that the Hamiltonian $H$ may contain multiple terms that act on the same subset, i.e., the subsets $C_i$ might not all be distinct.  

The string ``$1^s$'' is simply a device to ensure that the gap between the ``YES'' and ``NO'' cases is not too small, relative to the ``size'' of the problem.  Intuitively, we are interested in instances where $k$ is a constant, $m \leq \poly(n)$ and $s \leq \poly(n)$ (so $b-a \geq 1/\poly(n)$).  Then we can think of $n$ as the size of the problem, and we say an algorithm is efficient if it takes time $\poly(n)$.  

Some special cases of this problem are 2-Local Hamiltonian (where $k=2$), and 2-Local Hamiltonian on a graph $G$ (where $k=2$, and the graph $G'$, consisting of vertices $1,\ldots,n$ and edges $C_1,\ldots,C_m$, is restricted to be a subgraph of $G$).  In many cases the problem has been shown to be QMA-hard \cite{KSV,KKR,OT,qma1d-1,qma1d-2, NagajMozes06, kay, BiamonteLove07, SchuchVerstraete07}.

We remark that classical problems such as Max-k-SAT correspond to the special case of Local Hamiltonian where each $H_i$ is a diagonal matrix.

\vskipline

We define the Local Consistency problem as follows \cite{A}:  
\begin{quote}
Consider a system of $n$ qubits.  We are given a collection of local density matrices $\rho_1,\ldots,\rho_m$, where each $\rho_i$ acts on a subset of qubits $C_i \subseteq \set{1,\ldots,n}$ (and so has dimension $2^{|C_i|} \times 2^{|C_i|}$).  Each subset $C_i$ has size $|C_i| \leq k$, for some constant $k$.  

In addition, we are given a string ``$1^s$'' (the unary encoding of a natural number $s$), and a real number $\beta$, such that $\beta \geq 1/s$.  

All numbers are specified with $\poly(s)$ bits of precision.  

The problem is to distinguish between the following two cases:  
\begin{itemize}
\item There exists an $n$-qubit state $\sigma$ such that, for all $i$, $\Tr_{\set{1,\ldots,n}-C_i}(\sigma) = \rho_i$.  In this case, output ``YES.''  
\item For all $n$-qubit states $\sigma$, there exists some $i$ such that $\norm{\Tr_{\set{1,\ldots,n}-C_i}(\sigma) - \rho_i}_1 \geq \beta$.  In this case, output ``NO.''  
\end{itemize}
\end{quote}

Without loss of generality, we can assume that the subsets $C_i$ are all distinct; thus $m \leq \binom{n}{k} \leq n^k$.  As in the Local Hamiltonian problem, the string ``$1^s$'' is simply a device to ensure that the gap between the ``YES'' and ``NO'' cases is not too small, relative to the ``size'' of the problem.  Here, we use the norm $\norm{A}_1 = \Tr|A|$ to measure the distance between $\rho_i$ and the corresponding reduced density matrix of $\sigma$.  When multipled by 1/2, this is the trace distance.  

An important special case is where $k=2$, and we can visualize the system as a graph with nodes $1,\ldots,n$ and edges given by the subsets $C_1,\ldots,C_m$.  Local Consistency was shown to be QMA-hard in \cite{Liu-consistency-qma,Liu-thesis}, via an oracle reduction from Local Hamiltonian.  This reduction preserves the subsets $C_1,\ldots,C_m$; thus, for many of the special cases in which Local Hamiltonian is QMA-hard, Local Consistency is also QMA-hard.

We remark that when the matrices $\rho_i$ are all diagonal, this reduces to a classical problem of deciding the consistency of marginal probability distributions.

\vskipline

In sections 4 and 5 we will define special cases of Local Hamiltonian and Local Consistency, for 1-D and stoquastic systems.

\subsection{Convex Optimization using a Membership Oracle}

We review some algorithms for convex optimization using a membership oracle \cite{YN,GLS,BV,KV}, which were the main tool in reducing Local Hamiltonian to Local Consistency \cite{Liu-consistency-qma,Liu-thesis}.  This is a summary of the results in \cite{Liu-thesis}, which were based on \cite{GLS}.  

First, some notation:  let $S(p,r)$ denote the closed ball of radius $r$ around the point $p$, $S(p,r) = \set{x \in \RR^n \;|\; \norm{x-p} \leq r}$.  
Also, for any set $K$, we define the ball of radius $\varepsilon$ around $K$, 
\[
S(K,\varepsilon) = \set{x \in \RR^n \;|\; 
\text{there exists $y \in K$ s.t. $\norm{x-y} \leq \varepsilon$}}, 
\]
and we define the interior of $K$ with radius $\varepsilon$, 
\[
S(K,-\varepsilon) = \set{x \in \RR^n \;|\; S(x,\varepsilon) \subseteq K}.
\]

Let $K$ be a closed convex set in $\RR^n$, and suppose we are given a point $p \in \RR^n$, and inner and outer radii $r,R \in \RR$, such that $S(p,r) \subseteq K \subseteq S(0,R)$.  (This implies that $K$ is bounded and full-dimensional.)  We want to show a \textit{reduction} from the problem of optimizing a linear function over $K$, to the problem of deciding membership in $K$.  

We define the weak optimization problem $WOPT_\varepsilon$ as follows:  (The adjective ``weak'' refers to the fact that we allow additive errors of size $\varepsilon$.)  
\begin{verse}
Given $c \in \RR^n$, $\norm{c} = 1$, $\gamma \in \RR$, and $\varepsilon \in \RR$, $\varepsilon > 0$, all specified with $\poly(n)$ bits of precision.\\
If there exists a vector $y \in S(K,-\varepsilon)$ with $c \cdot y \geq \gamma + \varepsilon$, then answer ``YES.''\\
If for all $x \in S(K,\varepsilon)$, $c \cdot x \leq \gamma - \varepsilon$, then answer ``NO.''\\
\end{verse}

We define the weak membership problem $WMEM_\delta$ as follows:
\begin{verse}
Given $y \in \RR^n$, and $\delta \in \RR$, $\delta > 0$, all specified with $\poly(n)$ bits of precision.\\
If $y \in S(K,-\delta)$, then answer ``YES.''\\
If $y \notin S(K,\delta)$, then answer ``NO.''\\
\end{verse}


A reduction from $WOPT_\varepsilon$ to $WMEM_\delta$ is given in \cite{GLS}, using the shallow-cut ellipsoid method.  However, we need to modify their result slightly.  The reduction in \cite{GLS} uses a model for ``exact'' convex optimization, where $\varepsilon$ is exponentially small, and $\delta$ may be exponentially smaller than $\varepsilon$.  But for our application, $\varepsilon$ is inverse-polynomial in size, and we want $\delta$ to be at most polynomially smaller than $\varepsilon$; this can be described as ``approximate'' convex optimization.  

It turns out that the reduction of \cite{GLS} will suffice for our purposes, as long as $R/r$ is at most polynomially large.  The following statement is proved in \cite{Liu-thesis}.  

\begin{thm}
\label{thm-opt-mem}
Let $K$ be any closed convex set in $\RR^n$, such that $S(p,r) \subseteq K \subseteq S(0,R)$, as defined above.  Then there is an oracle reduction from $WOPT_\varepsilon$ to $WMEM_\delta$, for some $\delta \geq \poly(\varepsilon, (r/R), (1/n))$, which runs in time $\poly(n, (R/r), (1/\varepsilon))$.  
\end{thm}


We state a straightforward corollary of this result, that will be more convenient for our purposes.  First, we define a slightly modified problem, $WOPT^*_\varepsilon$, as follows:  
\begin{verse}
Given $c \in \RR^n$, $\norm{c} = 1$, $\gamma \in \RR$, and $\varepsilon \in \RR$, $\varepsilon > 0$, all specified with $\poly(n)$ bits of precision.\\
If there exists a vector $y \in K$ with $c \cdot y \geq \gamma + \varepsilon$, then answer ``YES.''\\
If for all $x \in K$, $c \cdot x \leq \gamma - \varepsilon$, then answer ``NO.''
\end{verse}

We also define a new problem, $WMEM^*_\delta$, as follows:  
\begin{verse}
Given $y \in \RR^n$, and $\delta \in \RR$, $\delta > 0$, all specified with $\poly(n)$ bits of precision.\\
If $y \in K$, then answer ``YES.''\\
If $y \notin S(K,\delta)$, then answer ``NO.''
\end{verse}

\begin{cor}
\label{cor-opt-mem}
Let $K$ be any closed convex set in $\RR^n$, such that $S(p,r) \subseteq K \subseteq S(0,R)$, as defined above.  Then there is an oracle reduction from $WOPT^*_\varepsilon$ to $WMEM^*_\delta$, for some $\delta \geq \poly(\varepsilon, (r/R), (1/n))$, which runs in time $\poly(n, (R/r), (1/\varepsilon))$.  
\end{cor}

\subsection{Reduction from Local Hamiltonian to Local Consistency}

We briefly review the main result of \cite{Liu-consistency-qma}.
\begin{thm}
\label{thm-QMA-hard}
There is a poly-time oracle reduction from Local Hamiltonian to Local Consistency.
\end{thm}

The main idea is that finding the lowest eigenvalue of a local Hamiltonian $H = H_1 + \cdots + H_m$ is equivalent to finding local density matrices $\rho_1,\ldots,\rho_m$ that minimize the linear function $f(\rho_1,\ldots,\rho_m) = \Tr(H_1\rho_1) + \cdots + \Tr(H_m\rho_m)$ over the convex set $K = \set{(\rho_1,\ldots,\rho_m) \text{ that are consistent}}$.  (Note that, if $\rho_1,\ldots,\rho_m$ are consistent with a global state $\sigma$, then $f(\rho_1,\ldots,\rho_m) = \Tr(H\sigma)$.)  Given an oracle for the Local Consistency problem, we can construct a membership oracle for the set $K$, then apply Corollary \ref{cor-opt-mem} to solve the optimization problem over $K$, and thus solve the Local Hamiltonian problem.

The main technical detail is to formulate the problem in such a way that $K$ has the necessary geometric properties, i.e., $K$ has outer radius $R$ and inner radius $r$ such that $R/r$ is at most polynomially large.  To this end, we will construct a set $\SSS$ of local observables; their expectation values encode the information contained in $\rho_1,\ldots,\rho_m$ without any redundancy.  

We introduce the $n$-qubit Pauli matrices $P = \Tensor_{i=1}^n P_i$, where $P_i \in \set{I,X,Y,Z}$.  Note that for any $n$-qubit Pauli matrices $P$ and $Q$, $\Tr(PQ) = 2^n$ if $P = Q$ and 0 otherwise.  Any $n$-qubit density matrix $\sigma$ can be written in the form 
\[
\sigma = \frac{1}{2^n} \sum_P \alpha_P P, \qquad \alpha_P = \Tr(P\sigma).
\]

Let $C$ be a subset of qubits.  We say that $P$ is supported inside $C$ if, for all $i \notin C$, $P_i = I$.  We claim that the expectation values $\alpha_P$, for those $P$ supported in $C$, contain precisely the same information as the reduced density matrix on $C$.  To see this, define $P|_C = \Tensor_{i \in C} P_i$, which we call the ``restriction'' of $P$ to $C$.  Then we can write the reduced density matrix on $C$ in the form 
\[
\begin{split}
\Tr_{\set{1,\ldots,n}-C}(\sigma)
&= \frac{1}{2^n} \sum_{\text{$P$ supp. in $C$}} \alpha_P \Tr_{\set{1,\ldots,n}-C}(P)
 = \frac{1}{2^{|C|}} \sum_{\text{$P$ supp. in $C$}} \alpha_P P|_C.
\end{split}
\]

For a collection of subsets $C_1,\ldots,C_m$, we define $\SSS$ to be the set of ``local'' Pauli matrices, excluding the identity matrix, 
\[
\SSS = \bigcup_{i=1}^m \set{P \;|\; \text{$P$ is supported inside $C_i$}} - \set{I}.
\]
We also let $D = |\SSS|$, and note that $D \leq 4^k m - 1$.  We can now replace the local density matrices $\rho_1,\ldots,\rho_m$ with the set of expectation values $\alpha_P$ for the observables $P \in \SSS$.  Using this formulation, the convex set $K$ has the required geometric properties.  (This uses the orthogonality properties of $P$; see \cite{Liu-consistency-qma} for details.)

\section{Reduction from Local Consistency to Local Hamiltonian}
\label{ch4-consistency-localham}


The idea comes from a theorem of ``strong alternatives'' in semidefinite programming \cite{BV}.  Let $F_1,\ldots,F_D$ be complex Hermitian matrices of dimension $N$.  Consider the following matrix inequality:
\begin{equation}
\label{ch4-eqn-a}
\sum_{i=1}^D x_i F_i + I \prec 0, 
\end{equation}
where $x \in \RR^D$ is a variable.  (Notation:  $M \prec 0$ means $M$ is strictly negative definite, $M \succeq 0$ means $M$ is positive semidefinite, etc.)  Also consider the following system of inequalities:
\begin{equation}
\label{ch4-eqn-b}
Z \succeq 0, \quad Z \neq 0, \quad \Tr(F_i Z) = 0 \; (\forall i = 1,\ldots,D), 
\end{equation}
where $Z$, a complex Hermitian matrix of dimension $N$, is a variable.  The theorem states that exactly one of the two inequalities (\ref{ch4-eqn-a}) and (\ref{ch4-eqn-b}) is feasible.  In other words, if (\ref{ch4-eqn-b}) is feasible, then (\ref{ch4-eqn-a}) is not; and if (\ref{ch4-eqn-b}) is not feasible, then (\ref{ch4-eqn-a}) is.  (When this property holds, we say that (\ref{ch4-eqn-a}) and (\ref{ch4-eqn-b}) are strong alternatives.)

Observe that inequality (\ref{ch4-eqn-b}) can be used to express the Local Consistency problem:  $Z$ is a global density matrix (unnormalized, but note that all the constraints remain the same if we divide across by $\Tr(Z)$), and we can choose the constraints $\Tr(F_i Z) = 0$ to ensure that $Z$ agrees with the desired local density matrices (note that the matrices $F_i$ will then be local observables).  But now the expression $\sum_{i=1}^D x_i F_i + I$ in inequality (\ref{ch4-eqn-a}) is simply a local Hamiltonian, and estimating its largest eigenvalue is precisely the Local Hamiltonian problem (modulo a sign flip).  So a Local Hamiltonian oracle allows us to test membership in the convex set defined by inequality (\ref{ch4-eqn-a}); and, using the methods of convex optimization described in Chapter 2, we can then decide the feasibility of (\ref{ch4-eqn-a}).  Since (\ref{ch4-eqn-a}) and (\ref{ch4-eqn-b}) are strong alternatives, this lets us solve the Local Consistency problem.

This is the intuition, but some further work is needed to make it rigorous.  We have to allow for the inverse-polynomial precision in the Local Consistency and Local Hamiltonian problems.  Also, in order to do convex optimization with a membership oracle, the set of feasible solutions $K$ must satisfy certain geometric properties.  (In particular, $K$ must be finite!)  So we have to formulate inequality (\ref{ch4-eqn-a}) in a different way.  We will show a finite-precision, ``algorithmic'' version of the theorem of strong alternatives.

\begin{thm}
\label{ch4-thm-consistency-localham}
There is a poly-time oracle reduction from Local Consistency to Local Hamiltonian.
\end{thm}

\noindent
Proof:  Suppose we have an instance of the Local Consistency problem, i.e., a collection of local density matrices $\rho_1,\ldots,\rho_m$, where $\rho_i$ describes a subset of qubits $C_i$, and an error parameter $\beta$.  Let $\SSS$ be the set of local Pauli observables, and let $D = |\SSS|$.  For each observable $P \in \SSS$, we define $\alpha_P$ to be the desired expectation value, which we compute as follows:  pick some subset $C_i$ such that $P$ is supported in $C_i$, then set $\alpha_P = \Tr(P\rho_i)$.  

We write down a convex program and its dual.  For each $P \in \SSS$, we define a new observable 
\[
F_P = P - \alpha_P I, 
\]
which is shifted so that the desired expectation value now equals 0.  We also define $F(x)$ to be a linear combination of these observables, 
\[
F(x) = \sum_{P \in \SSS} x_P F_P + I, \quad \text{for $x \in \RR^D$}.  
\]

Now consider the following convex program:  
\begin{verse}
Find some $x \in [-1,1]^D$ and $s \in [1-2D,1+2D]$ that \\
minimize $s$ such that $F(x) \preceq sI$.
\end{verse}
To see that this is a convex program, recall that the largest eigenvalue of $F(x)$ is a convex function of $x$, since it can be written as the pointwise minimum over a family of affine functions of $x$.  The variable $s$ is redundant here, but it will play a role later when we apply algorithms to solve this program.  We will refer to this as the \textit{primal program}; let $p^*$ denote the optimal value of the objective function $s$.

The \textit{dual program} is as follows:
\begin{verse}
Find some $2^n \times 2^n$ complex matrix $Z$ that \\
maximizes $g(Z)$ such that $Z \succeq 0$ and $\Tr(Z) = 1$,
\end{verse}
where the dual function $g(Z)$ is given by 
\[
\begin{split}
g(Z) &= \inf_{\substack{x \in [-1,1]^D \\ s \in [1-2D,1+2D]}} s + \Tr(Z(F(x)-sI)) \\
     &= \inf_{x \in [-1,1]^D} \Tr(ZF(x)) \\
     &= \inf_{x \in [-1,1]^D} \sum_{P \in \SSS} x_P \Tr(ZF_P) + 1.
\end{split}
\]
Let $d^*$ denote the optimal value of the objective function $g(Z)$.  Strong duality holds because the primal problem is convex and satisfies a generalized Slater condition \cite{BV} (to see this, note that the point $(x,s) = (0,2)$ is strictly feasible, i.e., it lies in the relative interior of the domain, and it satisfies $F(x) \prec sI$).  Strong duality implies that $p^* = d^*$, i.e., the optimal values of the primal and dual programs are equal.

\vskipline

We now give a poly-time oracle reduction from Local Consistency to Local Hamiltonian.  We show that Local Consistency reduces to the weak optimization problem $WOPT^*$, which reduces to the weak membership problem $WMEM^*$, which reduces to Local Hamiltonian.  

First, suppose we have a ``YES'' instance of Local Consistency.  Then there exists an $n$-qubit state $\sigma$ such that, for all $P \in \SSS$, $\Tr(P\sigma) = \alpha_P$.  So in the dual program there exists some $Z \succeq 0$, $\Tr(Z) = 1$, such that for all $P \in \SSS$, $\Tr(ZF_P) = 0$.  This implies $g(Z) = 1$, hence the dual program has optimal value $d^* \geq 1$.  By strong duality, the primal program has optimal value $p^* \geq 1$.  

On the other hand, suppose we have a ``NO'' instance of Local Consistency.  We claim that, for all $n$-qubit states $\sigma$, $\sum_{P \in \SSS} |\Tr(P\sigma) - \alpha_P| \geq \beta$.  This can be seen as follows.  Note that, for any $\sigma$, there is some subset $C_i$ such that $\norm{\tilde{\sigma}-\rho_i}_1 \geq \beta$, where $\tilde{\sigma} = \Tr_{\set{1,\ldots,n}-C_i}(\sigma)$.  Using the matrix Cauchy-Schwarz inequality \cite{Bhatia}, $\norm{\tilde{\sigma}-\rho_i}_1 \leq \norm{\tilde{\sigma}-\rho_i}_2 \sqrt{2^k}$.  By Fourier analysis, 
\[
\begin{split}
\norm{\tilde{\sigma}-\rho_i}_2
 &= \frac{1}{\sqrt{2^k}} \Bigl( \sum_\text{$P$ supp. on $C_i$} \Tr(P(\tilde{\sigma}-\rho_i))^2 \Bigr)^{1/2} \\
 &\leq \frac{1}{\sqrt{2^k}} \sum_{P \in \SSS} |\Tr(P(\tilde{\sigma}-\rho_i))|
 = \frac{1}{\sqrt{2^k}} \sum_{P \in \SSS} |\Tr(P\sigma) - \alpha_P|.  
\end{split}
\]
The claim follows by combining these inequalities.

Therefore, in the dual program, for all $Z$ such that $Z \succeq 0$ and $\Tr(Z) = 1$, we have that $\sum_{P \in \SSS} |\Tr(ZF_P)| \geq \beta$, which implies $g(Z) \leq 1-\beta$.  Thus the dual program has optimal value $d^* \leq 1-\beta$.  By strong duality, the primal program has optimal value $p^* \leq 1-\beta$.  

So we have reduced Local Consistency to the problem of distinguishing between the two cases $p^* \geq 1$ and $p^* \leq 1-\beta$ for the primal program.  This is an instance of the weak optimization problem $WOPT^*_{\beta/2}$ over the convex set 
\[
K = \set{(x,s) \in [-1,1]^D \times [1-2D,1+2D] \;|\; F(x)-sI \preceq 0}.
\]

\vskipline

Now we will reduce $WOPT^*$ to $WMEM^*$.  First we need some bounds on the geometry of $K$.  It is easy to see that $K$ is contained within a ball of radius $R = \sqrt{D+(1+2D)^2} \leq O(D)$.  In addition, we claim that $K$ contains a ball around the point $(0,\ldots,0,2)$ of radius $r = \frac{1}{4(D+1)}$.  To see this, consider an arbitrary point $(y,t+2)$ where $y \in \RR^D$, $t \in \RR$ and $\sqrt{\norm{y}^2+t^2} \leq \frac{1}{4(D+1)}$.  The operator 
\[
F(y) - (t+2)I = \sum_{P \in \SSS} y_P F_P - tI - I
\]
has all of its eigenvalues bounded above by $\sum_{P \in \SSS} \frac{1}{4(D+1)} \norm{F_P} + \frac{1}{4(D+1)} - 1 \leq -\frac{1}{2}$ (using the fact that $\norm{F_P} \leq 2$).  Thus $(y,t+2)$ is in $K$.  This proves the claim.  

So we have $R/r \leq O(D^2)$.  By Corollary \ref{cor-opt-mem}, $WOPT^*_{\beta/2}$ reduces to $WMEM^*_\delta$ where $\delta \geq \poly(\beta, 1/D)$, with running time $\poly(D,1/\beta)$.  

\vskipline

Finally, we reduce $WMEM^*$ to the Local Hamiltonian problem.  Observe that, since the $F_P$ are local operators, $F(x)$ is a local Hamiltonian.  Given an oracle that solves the Local Hamiltonian problem, we can estimate the largest eigenvalue of $F(x)$ (i.e., the smallest eigenvalue of $-F(x)$), and thus decide whether $(x,s)$ is in the set $K$.  

Suppose we have a ``YES'' instance of $WMEM^*_\delta$.  Then $(x,s) \in K$, so $F(x) \preceq sI$, i.e., all eigenvalues of $-F(x)$ are $\geq -s$.  So this is a ``NO'' instance of Local Hamiltonian.  

Now suppose we have a ``NO'' instance of $WMEM^*_\delta$.  Then $(x,s) \notin S(K,\delta)$, and in particular, $(x,s+\delta) \notin K$.  So $F(x) \npreceq (s+\delta)I$, i.e., $-F(x)$ has an eigenvalue that is $\leq -s-\delta$.  So this is a ``YES'' instance of Local Hamiltonian.  

Note that $\norm{F(x)} \leq \sum_{P \in \SSS} \norm{F_P} + 1 \leq 2D+1$.  Thus, $WMEM^*_\delta$ reduces to Local Hamiltonian with precision $\delta/(2D+1)$.  

Thus we conclude that Local Consistency (with precision $\beta$) reduces to Local Hamiltonian (with precision $\poly(\beta, 1/D)$), and the running time is $\poly(D, 1/\beta)$.  Note that $D < 4^km$ is polynomial in the size of the input.  $\square$

\section{Local Consistency for 1-D Systems}

Let us consider a 1-dimensional chain of $n$ qudits (a qudit is a $d$-dimensional particle), with nearest-neighbor interactions (i.e., interactions between particle $i$ and particle $i+1$, for $i = 1,\ldots,n-1$).  

First consider the case of qubits ($d=2$).  The reduction from Local Hamiltonian to Local Consistency shown in Theorem \ref{thm-QMA-hard}, and the reverse reduction shown in Theorem \ref{ch4-thm-consistency-localham}, both preserve the neighborhood structure of the problems---that is, each local term in the Hamiltonian corresponds to a local density matrix, and vice versa.  Thus we have:  
\begin{cor}
On a 1-D chain of qubits ($d=2$), Local Hamiltonian and Local Consistency have equivalent complexity (up to poly-time oracle reductions).  
\end{cor}

We will now sketch one way of extending these results to the case of qudits ($d>2$).  The first step is to define a set of observables for a single qudit, with nice properties similar to the Pauli matrices.  Let $\ket{i}$, $i = 0,\ldots,d-1$ denote the standard basis states for a single qudit.  Also, let $\text{i}$ (in plain, not italic type) denote the square root of $-1$.
\begin{align*}
X_{ij} &= \ket{j}\bra{i} + \ket{i}\bra{j}, \qquad 0 \leq i<j \leq d-1 \\
Y_{ij} &= \text{i}\ket{j}\bra{i} - \text{i}\ket{i}\bra{j}, \qquad 0 \leq i<j \leq d-1 \\
Z_i &= \Bigl( \frac{1}{i+1} \sum_{a=0}^i \ket{a}\bra{a} \Bigr) - \ket{i+1}\bra{i+1}, 
\qquad 0 \leq i \leq d-2
\end{align*}
Note that $Z_i$ is the diagonal matrix whose diagonal consists of $\frac{1}{i+1}$ in the first $i+1$ positions, followed by $-1$, followed by $0$ in all the remaining positions.  We have a total of $2\binom{d}{2} + (d-1) = d(d-1) + (d-1) = d^2 - 1$ observables.

These observables satisfy the following orthogonality relations:
\begin{center}
\begin{tabular}{|c|c|l|}
	\hline
	$A$ & $B$ & $\Tr(AB)$ \\
	\hline
	$I$ & $I$ & $d$ \\
	$I$ & $X_{kl}$ & 0 \\
	$I$ & $Y_{kl}$ & 0 \\
	$I$ & $Z_k$ & 0 \\
	\hline
	$X_{ij}$ & $X_{kl}$ & 2 if $(i,j)=(k,l)$; 0 otherwise \\
	$X_{ij}$ & $Y_{kl}$ & 0 \\
	$X_{ij}$ & $Z_k$ & 0 \\
	\hline
	$Y_{ij}$ & $Y_{kl}$ & 2 if $(i,j)=(k,l)$; 0 otherwise \\
	$Y_{ij}$ & $Z_k$ & 0 \\
	\hline
	$Z_i$ & $Z_k$ & $1+\frac{1}{i+1}$ if $i=k$; 0 otherwise \\
	\hline
\end{tabular}
\end{center}
In addition, note that $\norm{X_{ij}} = \norm{Y_{ij}} = \norm{Z_i} = 1$.

We can now use these qudit observables in the same way that we used the Pauli matrices for qubits.  We construct $n$-qudit observables by taking tensor products of single-qudit observables:  $P = \Tensor_{a=1}^n P_a$, where $P_a \in \set{I,X_{ij},Y_{ij},Z_i}$.  Note that for any $n$-qudit observables $P$ and $Q$, $\Tr(PQ) = \Tr(P^2)$ if $P = Q$ and 0 otherwise.  Any $n$-qudit density matrix $\sigma$ can be written in the form 
\[
\sigma = \sum_P \frac{\alpha_P}{\Tr(P^2)} P, \qquad \alpha_P = \Tr(P\sigma).
\]

As before, we say that $P$ is supported inside a subset $C \subseteq \set{1,\ldots,n}$ if for all $i \notin C$, $P_i = I$.  We define $P|_C = \Tensor_{i \in C} P_i$, the ``restriction'' of $P$ to the subset $C$.  We can write the reduced density matrix for the subset $C$ in the form 
\[
\begin{split}
\Tr_{\set{1,\ldots,n}-C}(\sigma)
&= \sum_{\text{$P$ supp. in $C$}} \frac{\alpha_P}{\Tr(P^2)} \Tr_{\set{1,\ldots,n}-C}(P)
 = \sum_{\text{$P$ supp. in $C$}} \frac{\alpha_P}{\Tr((P|_C)^2)} P|_C.
\end{split}
\]

Now we can use essentially the same reductions as before, from Local Hamiltonian to Local Consistency and vice versa, for systems of qudits.  (Details omitted.)  In particular, this implies:
\begin{cor}
On a 1-D chain of qudits (for any fixed $d \geq 2$), Local Hamiltonian and Local Consistency have equivalent complexity (up to poly-time oracle reductions).  
\end{cor}

\section{Stoquastic Local Consistency}

A Hamiltonian $H$ is called ``stoquastic'' if all of its off-diagonal matrix elements, relative to the standard basis, are less than or equal to 0.  (Note that the diagonal elements can be made $\leq 0$ by subtracting a multiple of the identity from $H$; this shifts the eigenvalues but does not change the eigenvectors.)  By the Perron-Frobenius theorem \cite{Bellman}, the ground state of $H$ has the form $\ket{\psi} = \sum_z c_z \ket{z}$ (up to an overall phase factor), where $\ket{z}$ are the standard basis vectors and $c_z \geq 0$.  (In other words, $\ket{\psi}$ is a superposition without any negative or complex coefficients; thus it resembles a classical probability distribution.)  

In the Stoquastic Local Hamiltonian problem, we are given a local Hamiltonian $H = H_1 + \cdots + H_m$ where each of the local terms $H_i$ is stoquastic.  As discussed previously, this makes the problem potentially easier \cite{stoq-1,stoq-2}.  In this section we propose a Stoquastic Local Consistency problem, and show that it has the same complexity as Stoquastic Local Hamiltonian (up to poly-time reductions).  

First, let us say that a density matrix $\rho$ is ``stoquastic'' if all of its off-diagonal matrix elements, relative to the standard basis, are greater than or equal to 0.  (Its diagonal elements must be $\geq 0$ since $\rho$ is positive semidefinite.)  Note that the set of stoquastic density matrices is convex.  

Now consider an obvious way of defining the Stoquastic Local Consistency problem:
\begin{quote}
Given local density matrices $\rho_1,\ldots,\rho_m$ which are stoquastic, does there exist a global density matrix $\sigma$ that agrees with $\rho_1,\ldots,\rho_m$?
\end{quote}
Stoquastic Local Hamiltonian easily reduces to this problem.  However, it is not clear whether this problem in turn reduces to Stoquastic Local Hamiltonian; the technique from section \ref{ch4-consistency-localham} instead produces a reduction from this problem to standard Local Hamiltonian.

Instead we will use a more subtle definition of Stoquastic Local Consistency:
\begin{quote}
Given local density matrices $\rho_1,\ldots,\rho_m$, does there exist a global density matrix $\sigma$ such that, for all $i=1,\ldots,m$, $\Tr_{\set{1,\ldots,n}-C_i}(\sigma) \geq_e \rho_i$?  

(Here, $C_i$ is the subset of qubits described by $\rho_i$, and $\geq_e$ denotes element-wise inequality between two matrices written in the standard basis; we assume all matrices are real.)
\end{quote}
This definition is a little unusual, but it turns out to have the desired property:  we can give a reduction from Stoquastic Local Hamiltonian to Stoquastic Consistency, and vice versa.  The element-wise inequality $\geq_e$ comes about naturally from the duality technique used in the second reduction.

We now state the full definition of the Stoquastic Local Consistency problem:
\begin{quote}
Consider a system of $n$ qubits.  We are given a collection of local density matrices $\rho_1,\ldots,\rho_m$, where each $\rho_i$ acts on a subset of qubits $C_i \subseteq \set{1,\ldots,n}$.  The matrices are assumed to be real (not complex).  Also, each subset $C_i$ has size $|C_i| \leq k$, for some constant $k$.  

In addition, we are given a string ``$1^s$'' (the unary encoding of a natural number $s$), and a real number $\beta$, such that $\beta \geq 1/s$. 

The problem is to distinguish between the following two cases:  
\begin{itemize}
\item There exists a real $n$-qubit state $\sigma$ such that, for all $i$, and for all $s,t \in \set{0,1}^{|C_i|}$, we have 
\[
\bra{s}\Tr_{\set{1,\ldots,n}-C_i}(\sigma)\ket{t} \geq \bra{s}\rho_i\ket{t}.  
\]
In this case, answer ``YES.''  
\item For all real $n$-qubit states $\sigma$, there exists some $i$, and there exist some $s,t \in \set{0,1}^{|C_i|}$, such that 
\[
\bra{s}\Tr_{\set{1,\ldots,n}-C_i}(\sigma)\ket{t} \leq \bra{s}\rho_i\ket{t} - \beta.  
\]
In this case, answer ``NO.''  
\end{itemize}
\end{quote}

We will show the following result:
\begin{thm}
Stoquastic Local Hamiltonian and Stoquastic Local Consistency have equivalent complexity (up to poly-time oracle reductions).
\end{thm}

\subsection{From Stoquastic Local Hamiltonian to Stoquastic Local Consistency}

First we show a reduction from Stoquastic Local Hamiltonian to Stoquastic Local Consistency.  The basic idea is as follows.  We are given a local Hamiltonian $H = \sum_{i=1}^m H_i$, where the $H_i$ are real and stoquastic.  Without loss of generality, we can assume $H_i \leq_e 0$ (we simply add a multiple of the identity to $H_i$).  Also, let $C_i$ be the subset of qubits on which $H_i$ acts.

Let $\rho_1,\ldots,\rho_m$ be local density matrices, where $\rho_i$ describes the subset of qubits $C_i$.  We want to find $\rho_1,\ldots,\rho_m$ that correspond to the ground state of $H$.  We consider the following convex program:
\begin{verse}
Find $\rho_1,\ldots,\rho_m$ that minimize $\sum_{i=1}^m \Tr(H_i\rho_i)$, 
subject to the constraints:
\begin{enumerate}
\item For all $i$, $\rho_i \succeq 0$ and $\Tr(\rho_i) = 1$.
\item There exists an $n$-qubit state $\sigma$ s.t. $\sigma \succeq 0$, $\Tr(\sigma) = 1$, \\
and for all $i$, $\Tr_{\set{1,\ldots,n}-C_i}(\sigma) \geq_e \rho_i$.
\end{enumerate}
Here, all matrices are restricted to be real.
\end{verse}

We claim that this convex program is equivalent to the Stoquastic Local Hamiltonian problem.  If $H$ has an eigenstate $\ket{\varphi}$ with eigenvalue $\leq \lambda$, then the convex program has optimal value $\leq \lambda$; to see this, set $\rho_i = \Tr_{\set{1,\ldots,n}-C_i} \ket{\varphi}\bra{\varphi}$.  On the other hand, if all the eigenvalues of $H$ are $\geq \lambda + \delta$, then the convex program has optimal value $\geq \lambda + \delta$; this follows because, for any feasible $\rho_1,\ldots,\rho_m$, we have that 
\[
\sum_{i=1}^m \Tr(H_i\rho_i) \geq \sum_{i=1}^m \Tr(H_i\sigma) = \Tr(H\sigma), 
\]
using constraint (2) and the fact that $H_i \leq_e 0$.  

Solving this convex program is an instance of the optimization problem $WOPT^*$ over the convex set $K$ defined by constraints (1) and (2).  By Corollary \ref{cor-opt-mem}, this reduces to the membership problem $WMEM^*$ over $K$.  This immediately reduces to Stoquastic Local Consistency.  

The main technical detail is to formulate the problem so that the set $K$ is full-dimensional, with inner and outer radii that satisfy $R/r \leq \poly(n)$.  This can be done using a subset of the local Pauli observables, as in Theorem \ref{thm-QMA-hard}, with two minor modifications.  

First, in this problem the matrices $\rho_1,\ldots,\rho_m$ are restricted to be real.  To represent these, we use the subset of local Pauli observables $P = \Tensor_{i=1}^n P_i$ that contain an even number of $Y$ factors.  This follows from the equation 
\[
\sigma = \frac{1}{2^n} \sum_P \alpha_P P, \qquad \alpha_P = \Tr(P\sigma); 
\]
note that $\alpha_P$ is always real (since $\sigma$ is Hermitian), while $P$ is real (imaginary) when it contains an even (odd) number of $Y$ factors.  One can check that there are $(4^n+2^n)/2$ real observables $P$, which correspond to the degrees of freedom for a $2^n \times 2^n$ real matrix.


Secondly, in this problem the set $K$ is larger than in Theorem \ref{thm-QMA-hard}, since the constraint (2) is weaker.  Nonetheless, constraint (1) ensures that $K$ is still contained within a ball of radius $R$.  $K$ also contains a ball of radius $r$, by the same argument as before.  


\subsection{From Stoquastic Local Consistency to Stoquastic Local Hamiltonian}

Next we show a reduction from Stoquastic Local Consistency to Stoquastic Local Hamiltonian.  The reduction uses strong duality, as in Section \ref{ch4-consistency-localham}.  

Suppose we have an instance of Stoquastic Local Consistency.  The first step is to represent $\rho_1,\ldots,\rho_m$ as the expectation values of certain observables.  However, we use a different set of observables, instead of the Pauli matrices, so that we can deal with inequalities involving the matrix elements of $\rho_i$.  These observables do not have any nice orthogonality properties, but the reduction technique from Section \ref{ch4-consistency-localham} does not require this.  

For each subset of qubits $C_i$, define the following observables acting on $C_i$:  
\[
X^{(i)}_{st} = \frac{1}{2}(\ket{s}\bra{t} + \ket{t}\bra{s}), 
  \qquad s,t \in \set{0,1}^{|C_i|}, \; s \preceq t, 
\]
where $s \preceq t$ denotes lexicographic order.  We can think of these observables as acting on the full $n$-qubit system (we tensor them with the identity matrix).  For any real $n$-qubit state $\sigma$, the matrix elements of $\Tr_{\set{1,\ldots,n}-C_i}(\sigma)$ are given by the expectation values of these observables:  
\[
\Tr(X^{(i)}_{st}\sigma) = \bra{s}\Tr_{\set{1,\ldots,n}-C_i}(\sigma)\ket{t}.  
\]
Then the conditions for a ``YES'' instance of Stoquastic Local Consistency can be written as:  
\[
\Tr(X^{(i)}_{st}\sigma) \geq \bra{s}\rho_i\ket{t}.  
\]
We let $\SSS$ be the set of all these observables $X^{(i)}_{st}$, for all of the subsets $C_i$, $i=1,\ldots,m$.  We also let $D = |\SSS|$.  Note that $\norm{X^{(i)}_{st}} \leq 1$.  

Next, we formulate a convex program, together with its dual.  Define new observables 
\[
F^{(i)}_{st} = X^{(i)}_{st} - \bra{s}\rho_i\ket{t} I, 
\]
which are shifted so that our goal is to satisfy the inequalities 
\[
\Tr(F^{(i)}_{st}\sigma) \geq 0.  
\]
For notational convenience, let us refer to these observables as $F_p$, for $p = 1,\ldots,D$.  Define $F(x)$ to be a linear combination of these observables, 
\[
F(x) = \sum_{p=1}^D x_p F_p + I, \quad \text{for $x \in \RR^D$}.  
\]

We construct a convex program which is similar to the one in Section \ref{ch4-consistency-localham}, except that we restrict $x$ to lie in the domain $[0,1]^D$ instead of $[-1,1]^D$.  (This restriction of the domain is the key feature that will eventually connect the dual program to Stoquastic Local Consistency.)
\begin{verse}
Find some $x \in [0,1]^D$ and $s \in [1-2D,1+2D]$ that \\
minimize $s$ such that $F(x) \preceq sI$.
\end{verse}
This is the \textit{primal} program; let $p^*$ denote the optimal value of the objective function $s$.

The \textit{dual} program is as follows:
\begin{verse}
Find some $2^n \times 2^n$ real matrix $Z$ that \\
maximizes $g(Z)$ such that $Z \succeq 0$ and $\Tr(Z) = 1$,
\end{verse}
where the dual function $g(Z)$ is given by 
\[
g(Z) = \inf_{x \in [0,1]^D} \Tr(ZF(x)) = \inf_{x \in [0,1]^D} \sum_{p=1}^D x_p \Tr(ZF_p) + 1.
\]
Let $d^*$ denote the optimal value of the objective function $g(Z)$.  Strong duality holds because the primal problem is convex and satisfies a generalized Slater condition \cite{BV} (to see this, note that the point $(x,s) = ((1/3D)\vec{1},2)$ is strictly feasible).  Strong duality implies that $p^* = d^*$.

Now, suppose we have a ``YES'' instance of Stoquastic Local Consistency.  Then in the dual program there exists some $Z \succeq 0$, $\Tr(Z) = 1$, such that for all $p$, $\Tr(ZF_p) \geq 0$.  This implies $g(Z) = 1$, hence the dual program has optimal value $d^* \geq 1$.  By strong duality, the primal program has optimal value $p^* \geq 1$.  

On the other hand, suppose we have a ``NO'' instance of Stoquastic Local Consistency.  Then for all $Z$ such that $Z \succeq 0$ and $\Tr(Z) = 1$, there is some $p$ such that $\Tr(ZF_p) \leq -\beta$, which implies $g(Z) \leq 1-\beta$.  Thus the dual program has optimal value $d^* \leq 1-\beta$.  By strong duality, the primal program has optimal value $p^* \leq 1-\beta$.  

Thus we have reduced Stoquastic Local Consistency to the primal problem.  This is an instance of the optimization problem $WOPT^*$ over the set 
\[
K = \set{(x,s) \in [0,1]^D \times [1-2D,1+2D] \text{ such that } F(x) \preceq sI}.
\]
Using a similar analysis as in section \ref{ch4-consistency-localham}, we can show that $K$ has inner and outer radii that satisfy $R/r \leq \poly(D)$.  (For instance, one can center the inner ball around the point $((1/3D)\vec{1},2)$.)  By Corollary \ref{cor-opt-mem}, this problem then reduces to the membership problem $WMEM^*$ for $K$.  

We claim that we can solve this problem, given an oracle for Stoquastic Local Hamiltonian.  Observe that the $F_p$ are local operators, whose off-diagonal elements are all $\geq 0$.  Thus $-F(x)$ is a stoquastic local Hamiltonian, and we can use the oracle to estimate its ground state energy.  This is equivalent to estimating the largest eigenvalue of $F(x)$, which allows us to test whether the constraint $F(x) \preceq sI$ is satisfied, and thus test membership in $K$.  


\section{Discussion}


We have shown that Local Hamiltonian and Local Consistency have equivalent complexity in two special cases, 1-D chains of qudits with small $d$ and stoquastic quantum systems, where neither problem is known to be QMA-hard.  To do this, we used ideas from \cite{Liu-consistency-qma}, together with a new reduction from Local Consistency to Local Hamiltonian using semidefinite programming duality.

In practice, one would like to solve particular instances of Local Hamiltonian that arise in condensed matter physics or quantum chemistry.  Reducing the problem to Local Consistency is one possible approach.  (In fact, this is the underlying idea in 2-RDM methods in quantum chemistry \cite{MazziottiBook07}.)  However, since Local Consistency is QMA-hard, it might seem that this only leads to a more general (hence potentially harder) problem.  Our results show that this is not always the case:  for 1-D and stoquastic systems (without any other special features), an efficient solution to Local Hamiltonian \textit{implies} an efficient solution to Local Consistency.  It would be interesting to find more instances where Local Consistency is a potentially tractable problem.

Finally, one might ask whether Stoquastic Local Consistency sheds any light on the class StoqMA introduced in \cite{stoq-2}.  (Recall that Stoquastic Local Hamiltonian is StoqMA-complete \cite{stoq-2}.)  Clearly, Stoquastic Local Consistency is StoqMA-hard, but we were not able to show that it is in StoqMA.  Could there be another complexity class that better describes these problems?

\vskipline

\noindent
\textit{Acknowledgements:}  Thanks to Frank Verstraete and Daniel Nagaj for useful discussions.  Part of this work was done while the author was a graduate student at the University of California, San Diego.  Supported by an ARO/DTO Quantum Computing Graduate Fellowship, and an NSF Mathematical Sciences Postdoctoral Fellowship.



\bibliographystyle{plain}
\bibliography{thesis}

\end{document}